\documentclass{jfm}
\usepackage{mathtools}
\usepackage{color}
\usepackage{graphicx}
\usepackage{natbib}
\usepackage{bm}
\usepackage{comment}
\usepackage{amsmath}
\definecolor{CG}{rgb}{0.2,0.8,0.2}
\ifCUPmtlplainloaded \else
  \checkfont{eurm10}
  \iffontfound
    \IfFileExists{upmath.sty}
      {\typeout{^^JFound AMS Euler Roman fonts on the system,
                   using the 'upmath' package.^^J}%
       \usepackage{upmath}}
      {\typeout{^^JFound AMS Euler Roman fonts on the system, but you
                   dont seem to have the}%
       \typeout{'upmath' package installed. JFM.cls can take advantage
                 of these fonts,^^Jif you use 'upmath' package.^^J}%
      }
  \else
  \fi
\fi

\ifCUPmtlplainloaded \else
  \checkfont{msam10}
  \iffontfound
    \IfFileExists{amssymb.sty}
      {\typeout{^^JFound AMS Symbol fonts on the system, using the
                'amssymb' package.^^J}%
       \usepackage{amssymb}%
       \let\le=\leqslant  
       \let\ge=\geqslant  
      }{}
  \fi
\fi

\ifCUPmtlplainloaded \else
  \IfFileExists{amsbsy.sty}
    {\typeout{^^JFound the 'amsbsy' package on the system, using it.^^J}%
     \usepackage{amsbsy}}
    {\providecommand\boldsymbol[1]{\mbox{\boldmath $##1$}}}
\fi

\newsavebox{\astrutbox}
\sbox{\astrutbox}{\rule[-5pt]{0pt}{20pt}}

\title{Modeling size segregation of granular materials: the roles of segregation, advection and diffusion}

\author[Yi Fan, Conor P. Schlick, Paul B. Umbanhowar, Julio M. Ottino and Richard M. Lueptow]%
{Y\ls I\ns F\ls A\ls N$^{1,5}$
  \thanks{Email address: yfan5@dow.com},\ns
C\ls O\ls N\ls O\ls R\ns P.\ns S\ls C\ls H\ls L\ls I\ls C\ls K$^2$,\ns \break
P\ls A\ls U\ls L\ns B.\ns U\ls M\ls B\ls A\ls N\ls H\ls O\ls W\ls A\ls R$^1$,\ns \break
J\ls U\ls L\ls I\ls O\ns M.\ns O\ls T\ls T\ls I\ls N\ls O$^{1,3,4}$\break
\and R\ls I\ls C\ls H\ls A\ls R\ls D\ns M.\ns L\ls U\ls E\ls P\ls T\ls O\ls W$^{1,4}$\ns\thanks{Email address: r-lueptow@northwestern.edu}}
\shortauthor{Fan,~Schlick,~Umbanhowar,~Ottino,~and~Lueptow}
\shorttitle{Modeling size segregation of granular materials}
\affiliation{$^1$Department of Mechanical Engineering, Northwestern University,
Evanston, IL 60208, USA\\[\affilskip]
$^2$Department of Engineering Science and Applied Mathematics, Northwestern University,
Evanston, IL 60208, USA\\
$^3$Department of Chemical and Biological Engineering, Northwestern University,
Evanston, IL 60208, USA\\
$^4$The Northwestern University Institute on Complex Systems (NICO), Northwestern University, Evanston, IL 60208, USA\\
$^5$The Dow Chemical Company, Midland, MI 48667, USA}
\date{?; revised ?; accepted ?. - To be entered by editorial office}
\begin{document}

\maketitle

\begin{abstract}
Predicting segregation of granular materials composed of different-sized particles is a challenging problem. In this paper, we develop and implement a theoretical model that captures the interplay between advection, segregation, and diffusion in size bidisperse granular materials. The fluxes associated with these three driving factors depend on the underlying kinematics, whose characteristics play key roles in determining particle segregation configurations. Unlike previous models for segregation, our model uses parameters based on kinematic measures from discrete element method simulations instead of arbitrarily adjustable fitting parameters, and it achieves excellent quantitative agreement with both experimental and simulation results when applied to quasi-two-dimensional bounded heaps. The model yields two dimensionless control parameters, both of which are only functions of physically control parameters (feed rate, particle sizes, and system size) and kinematic parameters (diffusion coefficient, flowing layer depth, and percolation velocity). The P\'eclet number, $Pe$, captures the interplay of advection and diffusion, and the second dimensionless parameter, $\Lambda$, describes the interplay between segregation and advection. A parametric study of $\Lambda$ and $Pe$ demonstrates how the particle segregation configuration depends on the interplay of advection, segregation, and diffusion. The model can be readily adapted to other flow geometries.
\end{abstract}

\begin{keywords}
Authors should not enter keywords on the manuscript, as these must be chosen by the author during the online submission process and will then be added during the typesetting process (see http://journals.cambridge.org/data/\linebreak[3]relatedlink/jfm-\linebreak[3]keywords.pdf for the full list)
\end{keywords}

\section{Introduction}
\label{intro}

Mixtures of granular material composed of particles with different sizes, densities, or other material properties, exhibit a propensity to segregate when subject to external excitation such as vibration \citep{Rosato1987,Knight1993,Kudrolli2004} or shear \citep{Savage1988, Ottino2000, Meier2007}. In particular, sheared granular mixtures differing in particle size present a common and challenging problem in many industrial contexts due to their tendency to form undesirable inhomogeneous particle configurations in tumblers, heaps, chutes, and silos \citep{pouliquen1997,Makse1997,Ottino2000,Aranson2006,Meier2007,Fan2012,Bridgwater2012}. These sorted (or unmixed) configurations resulting from size segregation occur also in natural phenomena such as debris flows \citep{Iverson1997}.

Many studies have been devoted to understanding the underlying mechanisms and developing predictive frameworks for size segregation and pattern formation in polydisperse, sheared granular flow \citep{Drahun1983,Ottino2000,yoon2006,Meier2007,fan2010,Fan2011PRL,Fan2011NJP,Christov11}. These studies have identified several driving mechanisms for segregation, pattern formation, and mixing of bidisperse particles. In the dilute, energetic flow regime, where particles interact mainly through binary collisions, the gradient of granular temperature alone can drive size segregation, which is successfully modeled by kinetic theory \citep{jenkins1989,hsiau1996,khakhar1999,arnarson1998,GALVIN2005,yoon2006}. In contrast, in the dense granular flow regime, particle geometry appears to be the primary driving mechanism. A percolation mechanism \citep{Williams1968,Drahun1983,Savage1988,Ottino2000} in which smaller particles are more likely than larger particles to fall through shear generated voids results in smaller particles moving downward while larger particles move upward, an effect that is characterized by a ``percolation'' velocity. Segregation due to percolation, along with other effects including advection \citep{Hill1999}, convection by secondary flow \citep{Khosropour1997,fan2010}, and collisional diffusion \citep{Hill1999,khakhar1999,Gray2006}, determine particle distributions in bidisperse dense flow, sometimes leading to complex patterns \citep{Meier2007,Christov11}.

Recently \citet{Larcher2013} extended kinetic theory to predict segregation of binary granular mixtures in dense flow and obtained qualitative agreement with discrete element method (DEM) simulations, but a first-principles based theory capable of quantitatively predicting segregation for dense flow is still lacking. In the meantime, a broad theoretical framework for segregation-driven pattern formation in dense flows is emerging \citep{Gray2005,Gray2006,thornton2006,May2010,Wiederseiner2011,Marks2011,thornton2012,kowalski2013}. This framework incorporates advection due to mean flow, percolation-driven segregation, and diffusion due to random particle collisions, in a continuum transport equation for the volume concentration of species $i$:
\begin{equation}
\frac {\partial c_i}{\partial t}= - \nabla\cdot{({\pmb u_i}c_i)} + \nabla\cdot(D\nabla c_i),
\label{transport}
\end{equation}
where ${\pmb u_i}$ is the velocity of species $i$ and $D$ is the collisional diffusion coefficient. Although diffusion can be anisotropic \citep{Utter2004}, for simplicity here we assume that $D$ is isotropic. This approach has yielded results that qualitatively reproduce data from experiment and simulation in a variety of flows including plug \citep{Gray2006}, chute \citep{Marks2011,Wiederseiner2011,thornton2012}, and annular shear \citep{May2010}. However, quantitative agreement has, until this work, been harder to achieve. Possible reasons for the lack of quantitative agreement in earlier work are the omission of one or more of advection, diffusion, or the dependence of percolation velocity on spatially varying shear rate \citep{Drahun1983}.

\begin{figure}
  \centerline{\includegraphics[scale=0.32]{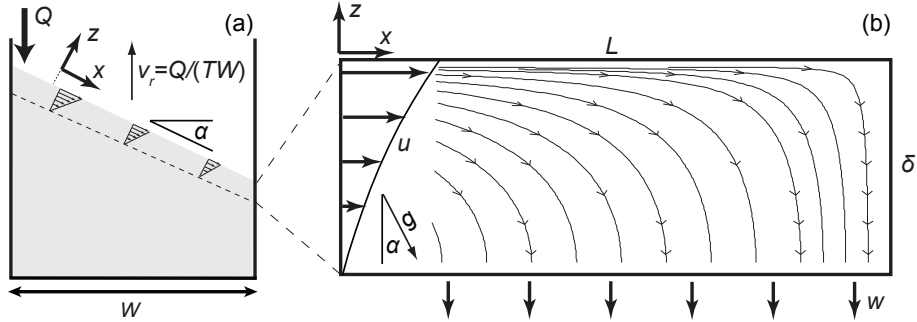}}% Images in 100% size
  \caption{(a) Sketch of a quasi-2D bounded heap of width $W$ with dynamic repose angle, $\alpha$. Granular material is fed onto the heap at a volumetric feed rate $Q$, and the heap rises with velocity $v_r=Q/TW$, where $T$ is the gap thickness between the front and back walls. (b) Sketch (not to scale) of the flowing layer in a rotated coordinate system (see text). We model the entire flowing layer from the granular material inlet to the downstream end of the flowing layer where the flow vanishes. Particles enter the flowing layer at the left boundary (the inlet) after leaving the feed zone and exit the flowing layer at the bottom boundary at uniform normal velocity. $\delta$ and $L$ are the thickness and length of the flowing layer, respectively. $u$ and $w$ are velocity components in the $x-$ and $z-$directions, respectively.}
\label{geometry}
\end{figure}

In this work we include all three mechanisms and examine their effects on segregation in a granular flow with non-trivial spatial variation: the quasi-two-dimensional (quasi-2D) bounded heap (figure \ref{geometry}(a)). In quasi-2D bounded heap flow, granular material is fed by gravity at one end of the heap, flows down hill in a thin flowing layer at the free surface, and is ultimately constrained by the downstream endwall. In the steady filling stage (after the heap reaches the downstream bounding endwall) for continuous (non-avalanching) flow, the free surface rises steadily and uniformly along the length of the heap with rise velocity $v_r$, and the local flow rate decreases linearly along the streamwise direction to zero at the downstream endwall. We use a coordinate system rising at $v_r$ with its $z-$axis rotated clockwise from vertical by the dynamic angle of repose, $\alpha$, where $x$ is the streamwise direction, $y$ is the spanwise direction, $z$ is the free surface normal direction, and the origin is at the intersection of the free surface and the right edge of the inlet feed stream (dashed line extending from the $z-$axis in (a)). We denote the velocity components as ($u,v,w$) in the $x-, y-$, and $z-$direction, respectively. The decrease of local flow rate induces a gradient of $u$ in the $x-$direction. In addition, $u$ and $w$ decrease from maximum values at the free surface to zero at the bottom of the flowing layer in the $z-$direction \citep{Fan2013}.

When a bidisperse mixture falls onto a heap and flows downhill, small particles fall into voids between large particles and sink to the bottom of the flowing layer, while large particles rise to the free surface. As the heap rises, small particles drop out of the flowing layer sooner and remain in the upstream region of the heap, while large particles are advected to the downstream region of the heap. This results in a separation of large and small particles in the streamwise direction \citep{Williams1963,Williams1968,Shinohara1972,Drahun1983,Goyal2006,Fan2012}, which is also similar to segregation patterns in avalanche flows \citep{pouliquen1997,Gray2009}.

There have been several attempts at modeling size segregation in bounded heap flow. \citet{Shinohara1972} proposed a screening layer model in which the flowing layer is divided into three sublayers -- large particles, mixed particles, and small particles -- from the free surface to the bottom of the flowing layer. Invoking mass conservation in each sublayer and accounting for particle migration between adjacent layers, they derived a model that qualitatively predicted the local particle concentration of each species with several arbitrarily adjustable fitting parameters (e.g.\ the velocity ratio of different sublayers and the penetration rate of segregating components in \citet{Rahman2011}). Boutreux \textit{et al.} \citep{Boutreux1996,boutreux1998} modeled particle exchange between the flowing layer and the static bed using a set of collision functions that are \textit{a priori} unknown and incorporated these collision functions into the mass conservation equations of each species. This model predicts the local particle concentrations qualitatively, but requires fitting parameters without clear physical interpretation (e.g.\ the characteristic length of segregation in \citet{Goyal2006} from which Boutreux \textit{et al.}'s model was adopted). The necessity of fitting parameters and the lack of quantitative agreement with experiments in both models are likely due to the oversimplification of flow kinematics.

Here, we present a model for predicting local particle distributions in quasi-2D bounded heap flow using a general scalar transport equation~\eqref{transport} and incorporating the effects of kinematics through advection, segregation, and diffusion. The theoretical predictions match quantitatively with both experiments and simulations. Compared to the models by \citet{Shinohara1972} and Boutreux \textit{et al}. \citep{Boutreux1996,boutreux1998}, our model relies on parameters characterizing the kinematics of the flow instead of arbitrarily adjustable parameters. The model demonstrates that the particle configuration is determined by the interplay of advection, segregation, and diffusion, which can be characterized by two dimensionless parameters, $Pe$ and $\Lambda$, that depend only on physical on control parameters (feed rate, particle sizes, and system size) and kinematic parameters (diffusion coefficient, flowing layer depth, and percolation velocity), which have a clear physical meaning and are determined from DEM simulations. The framework developed here for bounded heap flow can be generalized to other flow geometries with non-trivial flow kinematics.

The paper is organized as follows. In \S \ref{simulation}, the transport equation for quasi-2D bounded heap flow is developed and DEM simulations are used to obtain the values and expressions used in the model, including the mean velocity profiles, the percolation velocity, and the diffusion coefficient. In \S \ref{govern}, a dimensionless governing equation is developed, which yields the two dimensionless control parameters, $Pe$ and $\Lambda$. The theoretical predictions are compared with both DEM simulations and experiments in \S \ref{results}, and a systematic parametric study is carried out in \S \ref{discussion} to elucidate how $Pe$ and $\Lambda$ control particle configurations in the heap through the underlying physics. Concluding remarks are provided in \S \ref{conclusion}.

\section{Flow modeling and characterization}
\label{simulation}

\subsection{Transport equation for bounded heap flow}

To model segregation of bidisperse granular material in bounded heap flow, we apply the transport equation to the flowing layer (see figure \ref{geometry}(b)) since this is where segregation occurs. A 2D moving reference frame is used, where $x$ and $z$ denote the streamwise and normal directions, respectively, and the origin is at the intersection of the free surface ($z=0$) and the rightmost edge of the inlet feed zone ($x=0$). Due to the quasi-2D nature of the flow, the 2D feed rate $q=Q/T$ is used to characterize the feed rate, where $T$ is the gap thickness between the two sidewalls and $Q$ is the volumetric feed rate. The flowing layer has thickness $\delta$. As shown in our recent study \citep{Fan2013}, $\delta$ remains nearly constant along most of the length of the flowing layer and decreases slightly at the downstream end of the flowing layer. Here, for simplicity, we assume $\delta$ is constant along the entire length of the flowing layer, so $\delta$ can be thought of as a measure of the feed rate (i.e. $\delta$ is a function of $q$). Later in this paper, we show that a constant $\delta$ produces spatial concentration profiles that match simulations and experiments.

For a binary mixture of different-sized particles, subscript $i$ denotes each species ($i=l$ and $i=s$ represent large particles and small particles, respectively). No subscript is used for variables describing the combined flow. The volume concentration of species $i$, $c_i$, is defined by $c_i=f_i/f$, where $f_i$ is the solids volume fraction of species $i$ and $f=\sum f_i$. As noted by \citet{Fan2013}, $f$ is nearly constant in the flowing layer of a bounded heap, so we assume $f$ is constant here.

For the quasi-2D bounded heap, we assume no net motion of species in the spanwise ($y$) direction ($v_i=0$), and the two species flow at the same velocity as the mean flow in the streamwise direction ($u_i=u$). The normal velocity component of species $i$ is written as $w_i=w+w_{p,i}$, where $w_{p,i}$ is normal component of the percolation velocity of species $i$ relative to the mean normal flow. The streamwise component of the percolation velocity is assumed to be negligible. With these assumptions, the transport equation~\eqref{transport} can be written as

\begin{equation}
\frac {\partial c_i}{\partial t}+ \underbrace{\frac {\partial (uc_i)}{\partial x} + \frac {\partial (wc_i)}{\partial z}}_{\text{advection}}+ \underbrace{\frac {\partial (w_{p,i}c_i)}{\partial z}}_{\text{segregation}}- \underbrace{\left[\frac {\partial }{\partial x} \left(D\frac {\partial c_i}{\partial x}\right)+\frac {\partial }{\partial z} \left(D\frac {\partial c_i}{\partial z}\right)\right]}_{\text{diffusion}} =0.
\label{transport_heap}
\end{equation}
The term $\partial (w_{p,i}c_i)/\partial z$ accounts for the transport of species $i$ due to percolation. In equation~\eqref{transport_heap} the local volume concentration of each species is determined by advection due to the mean flow, segregation due to percolation, and diffusion due to random particle collisions, similar to previous studies \citep{Gray2006,thornton2006,Wiederseiner2011,thornton2012}.

Equation~\eqref{transport_heap} can be solved with appropriate boundary conditions to obtain the local concentration of each species. However, to achieve this, the velocity profiles, the percolation velocity, and the diffusion coefficient are needed. In the rest of this section, results from DEM simulations are used to obtain values and expressions for these quantities.

\subsection{Simulation method and geometry}
\label{method}

\begin{figure}
  \centerline{\includegraphics[scale=0.7]{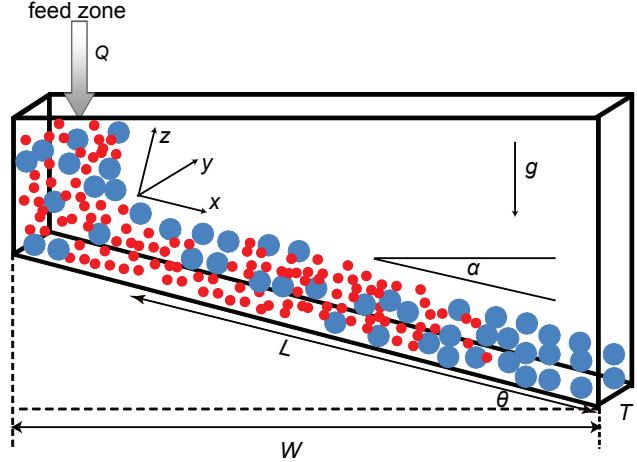}}% Images in 100% size
  \caption{Sketch (not to scale) of a quasi-2D bounded heap of width $W$ and thickness $T$ with the bottom wall inclined by angle $\theta$ used in DEM simulations. The effective length of the flowing layer from the right edge of the feed zone to the downstream bounding wall is $L$. }
\label{setup}
\end{figure}
In the DEM simulations, the translational and rotational motion of each particle are calculated by integrating Newton's second law. The forces between particles are repulsive and are non-zero only when particles are in contact. A linear-spring dashpot force model \citep{Cundall1979,Schafer1996,Ristow2000,chen2008} is used to calculate the normal force between two contacting particles:
\begin{equation}
\boldsymbol F_{ij}^n=[k_n\epsilon-2\gamma_n m_{{\rm eff}}(\boldsymbol V_{ij}\cdot\hat{\boldsymbol r}_{ij})]\hat{\boldsymbol r}_{ij}.
\end{equation}
Here, $\epsilon$ and $\boldsymbol V_{ij}= \boldsymbol V_i-\boldsymbol V_j$ denote the overlap and relative velocity of the two contacting particles $i$ and $j$, respectively. $\hat{\boldsymbol r}_{ij}$ is the unit vector between particles $i$ and $j$, and $m_{\rm{eff}}= (m_im_j)/(m_i+m_j)$ is the reduced mass. $k_n$ and $\gamma_n$ are stiffness and damping coefficients, respectively, and are related to the collision time $t_c$ and restitution coefficient $\varepsilon$ by $\gamma_n=-{\rm {ln} \varepsilon}/{t_c}$ and $k_n=[\left(\pi/t_c \right)^2+\gamma_n^2]m_{{\rm eff}}$ \citep{Schafer1996,Ristow2000}. For the tangential force, a linear spring model with Coulomb friction is used:
\begin{equation}
\boldsymbol F_{ij}^t=-{{\rm min}\left(|k_s \beta|, |\mu \boldsymbol F_{ij}^n| \right)}{\rm sgn}(\beta)\hat{\boldsymbol{s}}.
\end{equation}
The tangential displacement $\beta$ is given by $\beta(t)=\int_{t_s}^{t} \boldsymbol V_{ij}^s dt$ \citep{Rapaport2002}, where $t_s$ is the time of initial contact between two particles. $\boldsymbol V_{ij}^s$ is the relative tangential velocity of particles $i$ and $j$, and $\hat{\boldsymbol{s}}$ is the unit vector in the tangential direction. The tangential stiffness is $k_s= \frac{2}{7}k_n$ \citep{Schafer1996}. The velocity-Verlet algorithm \citep{Ristow2000} is used to update particle positions and velocities.

The bounded heap in these simulations is sketched in figure \ref{setup} and is identical in scale to our previous experiments \citep{Fan2012} and simulations \citep{Fan2013}. We simulate only the steady filling stage, which is similar to the experimental setup used by \citet{Drahun1983}. To do this, the bottom wall of the silo is inclined at an angle $\theta$ with respect to horizontal that is close to the dynamic angle of repose $\alpha$ in our previous experiments \citep{Fan2012}. During filling, particles that contact the inclined bottom wall are immobilized to increase the effective wall friction to prevent slip. When the heap is sufficiently deep ($\sim$10-15 particle diameters), the boundary effect of the bottom wall on the flowing layer is negligible, and the flow is comparable to the heap in experiments. For these simulations, the width of the silo $W$ is 45.7~cm and the gap thickness between the front and back walls $T$ is 1.27~cm. Particles are fed into the silo at the left end, 10~cm above the leftmost point of the bottom wall at volumetric flow rate $Q$.

Particles in the simulation have a material density $\rho=2500$~kg/m$^3$ and a restitution coefficient $\varepsilon=0.8$. Particle-particle and particle-wall friction coefficients, $\mu$, are 0.4. These values reflect those for spherical glass particles and have been confirmed in our previous study \citep{Fan2013}. To decrease computational time, the binary collision time is set to $t_c= 10^{-3}$~s, consistent with previous simulations \citep{chen2011,Fan2013} and sufficient for modeling hard spheres \citep{Silbert2007}. The integration time step is $t_c/100 = 1.0 \times 10^{-5}$~s to assure numerical stability. To reduce particle ordering, the diameter of each species is distributed uniformly between 0.9$d_i$ and 1.1$d_i$, where $d_i$ is the mean particle diameter for each species $i$. In the simulations, particle diameters range from 1 mm to 3 mm, and the size ratio varies from 1.5 to 3 (see table~\ref{table:percolation}). Up to one million particles are simulated. The DEM simulations have been validated in terms of flow kinematics and segregation by comparing with experiments \citep{Fan2013}.

\subsection{Mean velocity field}
\label{Kin_mean}
\begin{figure}
  \centerline{\includegraphics[scale=0.5]{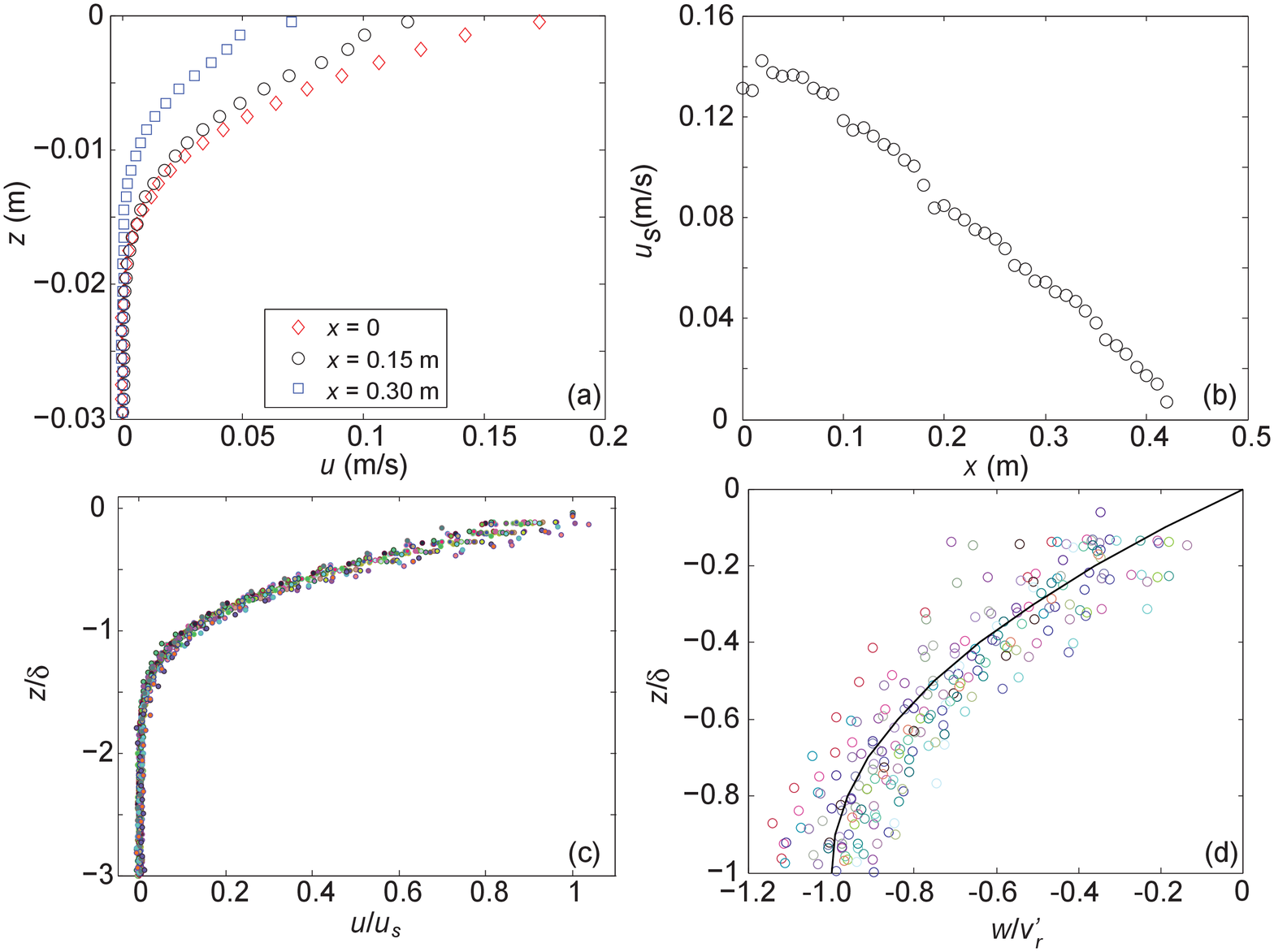}}% Images in 100% size
  \caption{Kinematics of a 1.5~mm and 3~mm diameter particle mixture at $Q=1.52\times 10^4$ ~mm$^3$/s. (a) Streamwise velocity profiles in the depth direction at three streamwise locations. (b) Surface velocity $u_s$ as a function of streamwise location $x$. (c) Scaled streamwise velocity profiles $u/u_s$ in the depth direction $z/\delta$ collapse onto a single curve at different streamwise locations. (d) Scaled normal velocity $w/v_r'$ in the depth direction at different streamwise locations (different colored symbols), where $v_r'=v_r{\rm cos}\alpha$ is the normal component of the rise velocity $v_r$. The curve represents the analytic solution from equation~\eqref{w_z_exp}. Results for mixtures of other particles are similar (see table \ref{table:percolation}). }
\label{kinematics}
\end{figure}
The mean velocity field was measured from the DEM simulations using the averaging method described in Appendix \ref{average}. Representative streamwise velocity profiles at different streamwise locations for one simulation (1.5 and 3~mm diameter particles, $Q=1.52\times 10^4$ ~mm$^3$/s) are shown in figure \ref{kinematics}(a). The streamwise velocity decreases rapidly from the free surface ($z=0$) in most of the flowing layer and then decays more slowly in the lower portion of the flowing layer ($-$0.01~m $\gtrsim z \gtrsim -0.015$~m) to the quasistatic region of the heap ($z < -0.015$~m). The streamwise velocity also decreases along the streamwise direction (figure \ref{kinematics}(a)). As shown in figure \ref{kinematics}(b), the streamwise velocity at the free surface, $u_s$, decreases linearly along the streamwise direction. The profiles of streamwise velocity at different streamwise locations collapse onto a single curve, as shown in figure \ref{kinematics}(c), when the streamwise velocity is normalized by the local surface velocity and $z$ is normalized by the local flowing layer thickness \citep {Fan2013}. This scaling is valid for different feed rates and particle size distributions, indicating that a universal functional form exists for the velocity field in bounded heap flow. Based on these results, the streamwise velocity in the flowing layer can be approximated as
\begin{equation}
u(x,z)=U\left(1-\frac{x}{L}\right)f(z),
  \label{u_x}
\end{equation}
where $f(z)$ characterizes the depth dependence with $f(0)=1$, and $U=u(0,0)$ is the velocity at the origin. The segregation model we consider allows any functional form $f(z)$ to characterize the velocity profile. Here we consider an exponential form $f(z)=e^{kz/\delta}$ and a linear form $f(z)=(1+z/\delta)$, both of which are reasonable approximations of the actual streamwise velocity profile \citep{Fan2013}.

Substituting equation~\eqref{u_x} into the mass conservation equation,
\begin{equation}
\frac {\partial u}{\partial x}+  \frac {\partial w}{\partial z}=0,
\label{mass_mix}
\end{equation}
integrating with the boundary condition $w=0$ at $z=0$ in the moving reference frame, and noting that $w$ is a function of $z$ only (uniform rise of the heap), an expression for the normal velocity $w(z)$ in the flowing layer is obtained:
%\begin{equation}
%  w(z)=\frac{1}{L}\int^0_{z} u(0,\xi)d\xi.
%  \label{w_z}
%\end{equation}
\begin{equation}
w(z)=-\frac{U}{L}\intop_{z}^{0}f(\xi)d\xi,
  \label{w_z}
\end{equation}
where $U$ is determined from $q$ and $f(z)$ as,
\begin{equation}
U=\frac{q}{\intop_{-\delta}^{0}f(\xi)d\xi}.
\label{u0}
\end{equation}

As shown in \citet{Fan2013}, an exponential expression $f(z)={e}^{kz/\delta}$ provides a reasonable approximation to the velocity profile (similar to previous results in other free surface flows including \citet{Komatsu2001} and \citet{Katsuragi2010}), where $k$ is a scaling constant. Combining equations~\eqref{u_x}, \eqref{w_z} and \eqref{u0}, an analytic expression for the mean velocity field is obtained:

\begin{equation}
  \label{w_z_exp}
  \begin{split}
   u&=\frac{kq}{\delta\left(1-e^{-k}\right)}\left(1-\frac{x}{L}\right)e^{kz/\delta}\\
   w&=\frac{q}{L\left(1-e^{-k}\right)}\left(e^{kz/\delta}-1\right).
  \end{split}
\end{equation}

Equation~\eqref{w_z_exp} automatically satisfies the boundary condition $w=-q/L=-v_r{\rm {cos}}\alpha$ at the bottom of the flowing layer ($z=-\delta$). In this study, we define the bottom of the flowing layer as the depth at which the streamwise velocity is 10\% of the surface velocity, which yields $k=2.3$. In our previous work \citep{Fan2013}, we found that a cut-off of 10\% matched other methods to determine the bottom of the flowing layer (e.\ g.\ \citet{GDRMidi2004} and \citet{Komatsu2001}). Different values of $k$ were tested (specifically $k=3$ and $k=4.6$, corresponding to cutoffs of 5\% and 1\%, respectively) for some of the results presented later in this paper, but $k=2.3$ produced the best match between experiment and simulation, as larger values include the ``creeping regime'', which is not considered in our model. Equation~\eqref{w_z_exp} for the normal velocity matches well the DEM simulation results, as shown in figure~\ref{kinematics}(d). The linear streamwise velocity profile $f(z)=(1+z/\delta)$ also provides a reasonable approximation \citep{GDRMidi2004,Socie2005,Fan2013}, and yields an alternative expression for the mean flow field:
\begin{equation}
  \label{w_z_linear}
  \begin{split}
  u&=\frac{2q}{\delta}\left(1-\frac{x}{L}\right)\left(1+\frac{z}{\delta}\right)\\
  w&=\frac{2q}{L}\left(\frac{z}{\delta}+\frac{z^{2}}{2\delta^{2}}\right).
  \end{split}
\end{equation}
We compare the accuracy of the linear and exponential velocity profiles in \S \ref{comparison}.

\subsection{Percolation velocity}
\label{Percolation}
\begin{figure}
  \centerline{\includegraphics[scale=0.5]{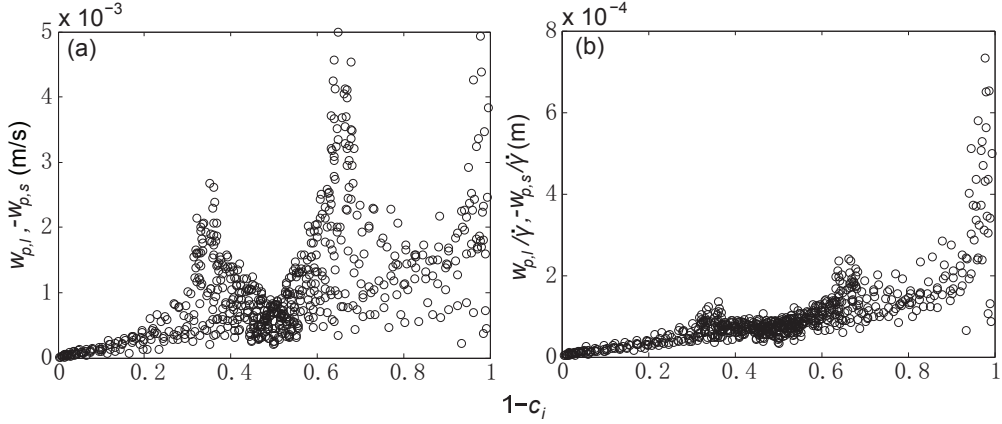}}% Images in 100% size
  \caption{Dependence of percolation velocity on particle concentration. (a) $w_{p,l}$ vs.\ $(1-c_l)$ and $-w_{p,s}$ vs.\ $(1-c_s)$. (b) $w_{p,l}/{\dot \gamma}$ vs.\ $(1-c_l)$ and $-w_{p,s}/\dot \gamma$ vs.\ $(1-c_s)$. Symbols denote over 500 data points spanning the entire length and depth of the flowing layer for a simulation run of 1 and 2~mm particles at $Q=1.52\times 10^4$ ~mm$^3$/s.}
\label{seg_v}
\end{figure}

The percolation velocity accounts for the relative motion between each species in the segregation direction. The percolation velocity depends on the particle size ratio, the strain rate, and the normal stress \citep{Bridgwater1978,Hill2008PRL,Golick2009}. However, in heap flow, since the flowing layer is only a few particle diameters thick ($<10d_l$) \citep {Fan2013}, the effect of the normal stress on percolation velocity can be safely neglected. Several models \citep{Shinohara1972,Savage1988,May2010,Marks2011} for the percolation velocity have been proposed, but none incorporated all of these parameters or were tested in different flow geometries.

Of course, when a granular mixture consists of different species with comparable volume fractions, the percolation velocity also depends on the local volume concentration of each species, since the void sizes are associated with the local packing. For example, percolation of a small particle will be enhanced when more large particles surround it. \citet{Savage1988} found that the percolation velocity of each species is proportional to the concentration of the other species, $w_{p,i} \thicksim (1-c_i)$. The same relation has been used in other studies \citep{dolgunin1998,Gray2006,Hajra2012}. Accordingly, the local percolation velocity of each species, $w_{p,i}$, and species volume concentration, $c_i$, are measured from DEM simulations, as described in Appendix \ref{average}, to investigate the dependence of percolation velocity on particle size ratio and shear rate in bounded heap flow. Figure~\ref{seg_v}(a) shows the percolation velocity (we plot the negative of the percolation velocity for small particles) of each species as a function of the local concentration of the other species for 1 and 2~mm particles at $Q=1.52\times 10^4$ ~mm$^3$/s in the flowing layer (over 500 data points are included spanning the entire length and depth of the flowing layer). To collapse the data in figure~\ref{seg_v}(a), the local percolation velocity is divided by the local shear rate, $\dot \gamma=\partial u/\partial z$, as shown in figure~\ref{seg_v}(b), since percolation only occurs when the material is dilated due to flow. The data over the entire length and depth of the flowing layer collapse and can be approximated by:
\begin{equation}
w_{p,l}=S\dot \gamma(1-c_l) \quad {\rm and} \quad w_{p,s}=-S\dot \gamma(1-c_s),
\label{percolation_v}
\end{equation}
where $S$, which has units of length, is the slope of the linear fit of the data in figure~\ref{seg_v}(b). $S$ represents the percolation length scale and depends both on the particle size ratio and absolute particle size. (Data for $1- c_i > 0.9$ were neglected when fitting a line to the data in figure~\ref{seg_v}(b), because the concentration of one species dominates the other species, and the fluxes of each species, $c_iw_{p,i}$, are much smaller than those for $1- c_i < 0.9$.) Equation~\eqref{percolation_v} satisfies mass conservation because the total net flux, $c_sw_{p,s}+c_lw_{p,l}$, is zero. As shown in table \ref{table:percolation}, the percolation length scale, $S$, is somewhat smaller than the size of the smallest particles. For the same particle mixture, $S$ is similar at different feed rates. At the same particle size ratio but different absolute particle sizes, $S$ is larger for the mixture of larger particles. At different size ratios, $S$ is difficult to compare, since the absolute size of the particles also plays a role in percolation.

\begin{table}
  \begin{center}
  \def~{\hphantom{0}}
  \begin{tabular}{ccccc}
      $R$ & $d_s$ (mm) &  $d_l$ (mm) & $Q$ (mm$^3$/s) & $S$ (mm)\\[3pt]
      1.5 & 1.0 & 1.5 & $1.52\times 10^4$  & 0.067 \\
      1.5 & 1.5 & 2.25 & $1.52\times 10^4$ & 0.2 \\
      1.5 & 2.0 & 3.0 & $1.52\times 10^4$ & 0.33 \\
      2.0 & 1.0 & 2.0 & $4.57\times 10^3$ & 0.19 \\
      2.0 & 1.0 & 2.0 & $1.52\times 10^4$ & 0.18 \\
      2.0 & 1.0 & 2.0 & $5.48\times 10^4$ & 0.17 \\
      2.0 & 1.5 & 3.0 & $1.52\times 10^4$ & 0.38 \\
      3.0 & 1.0 & 3.0 & $1.52\times 10^4$ & 0.29 \\
      3.0 & 1.0 & 3.0 & $5.48\times 10^4$ & 0.30 \\
  \end{tabular}
  \caption{Percolation length scale, $S$, for various particle mixtures and feed rates.}
  \label{table:percolation}
  \end{center}
\end{table}

\subsection{Diffusion}
\label{diffusion}

Granular material diffuses due to particle collisions, analogous in some ways to the Brownian motion of molecules and colloidal particles. For granular mixtures, the collisional diffusion, $D$, can result in re-mixing of segregating particles. Previous studies based on dimensional analysis \citep{bridgwater1980,campbell1997,savage1993}, experiments \citep{Utter2004}, and DEM simulations \citep{tripathi2013} have shown that in \textit{dense} granular systems of monodisperse particles or bidisperse particles differing only in material density, the diffusion coefficient scales as

\begin{equation}
D \thicksim \dot \gamma d^2.
\label{eq:diffusion}
\end{equation}
However, this relation has not been validated for dense granular flows of different-sized particles.

The local diffusion coefficient of the mixture in the segregation direction ($z-$direction) used here is determined by calculating the mean squared displacement, as described in Appendix \ref{average}. For simplicity of modeling, we use constant $D$, namely, the mean diffusion coefficient in the entire flowing layer measured directly from DEM simulations. The effect of using constant $D$ instead of a shear rate-dependent diffusion coefficient is evaluated in Appendix \ref{sensitivity}.

\section{Governing equation and numerical method}
\label{govern}

\subsection{Nondimensionalization}
\label{nondim}

Combining equation~\eqref{transport_heap} with equations~\eqref{mass_mix} and \eqref{percolation_v} yields the transport equation for species $i$,
\begin{equation}
\frac {\partial c_i}{\partial t}+  u\frac {\partial c_i}{\partial x} + w\frac {\partial c_i}{\partial z}+ S\frac {\partial }{\partial z}\left[{\dot \gamma} c_i(1-c_i)\right]- \left[\frac {\partial }{\partial x} \left(D\frac {\partial c_i}{\partial x}\right) + \frac {\partial }{\partial z} \left(D\frac {\partial c_i}{\partial z}\right)\right]=0.
\label{governing}
\end{equation}
In contrast to previous studies \citep{Gray2005,Gray2006,thornton2006,May2010,Wiederseiner2011,Marks2011,thornton2012}, when equation~\eqref{governing} is applied to bounded heap flow, it includes both the dependence of percolation velocity on spatially varying shear rate and the full effects of the kinematics on advection.

Equation~\eqref{governing} is nondimensionalized using
\begin{equation}
\tilde{x}=\frac {x}{L}, \quad \tilde{z}=\frac {z}{\delta}, \quad \tilde{t}=\frac {t}{\delta L/2q}, \quad \tilde{u}=\frac {u}{2q/\delta}, \quad {\rm and} \quad \tilde{w}=\frac {w}{2q/L}.
\label{normalization}
\end{equation}
In this way, the domain (the flowing layer) is transformed to a square ($0\le \tilde{x}\le 1$ and $-1\le \tilde{z}\le 0$), and the nondimensional governing equation for the concentration of species $i$ is:

\begin{equation}\label{eq:ND1}
\frac{\partial c_i}{\partial\tilde{t}}+\tilde{u}\frac{\partial c_i}{\partial\tilde{x}}+\tilde{w}\frac{\partial c_i}{\partial\tilde{z}}
+\Lambda(1-\tilde{x})\frac{\partial}{\partial\tilde{z}}\left[g(\tilde{z})c_i(1-c_i)\right]
=\left(\frac{\delta}{L}\right)^{2}\frac{\partial}{\partial\tilde{x}}\left(\frac{1}{{Pe}}\frac{\partial c_i}{\partial \tilde{x}}\right) +\frac{\partial}{\partial\tilde{z}}\left(\frac{1}{{Pe}}\frac{\partial c_i}{\partial \tilde{z}}\right),
\end{equation}
where $\Lambda=SL/\delta^{2}$, $Pe=2q\delta/DL$, and
\begin{equation}
g(\tilde{z})=\frac{1}{2}\frac{\delta f'(\delta\tilde{z})}{\intop_{-1}^{0}f(\delta\tilde{\xi})d\tilde{\xi}}.
\label{g}
\end{equation}
The dimensionless velocities ($\tilde u$ and $\tilde w$) and $g(\tilde z)$ are determined by equations~\eqref{w_z_exp} or \eqref{w_z_linear}, \eqref{normalization}, and \eqref{g}. The presence of $g(z)$ in the segregation term reflects the role of the functional form of the velocity profile, $f(z)$, as described further in \S \ref{comparison}. For $\delta/L\ll1$ ($\delta/L\approx1/50$ in our simulations), the diffusion term in the $x-$direction in equation~\eqref{eq:ND1} can be neglected, and thus the nondimensional governing equation becomes
\begin{equation}\label{eq:ND}
\frac{\partial c_i}{\partial\tilde{t}}+\tilde{u}\frac{\partial c_i}{\partial\tilde{x}}+\tilde{w}\frac{\partial c_i}{\partial\tilde{z}}
+\Lambda(1-\tilde{x})\frac{\partial}{\partial\tilde{z}}\left[g(\tilde{z})c_i(1-c_i)\right]
=\frac{\partial}{\partial\tilde{z}}\left(\frac{1}{{Pe}}\frac{\partial c_i}{\partial \tilde{z}}\right).
\end{equation}

The dimensionless parameters $\Lambda$ and $Pe$ in equation~\eqref{eq:ND} have clear physical meaning. $\Lambda$ is the ratio of an advection timescale, $L/u=L/(2q/\delta)$, to a segregation timescale, $\delta/w_p=\delta/(2Sq/\delta^2)$. $Pe$, the P\'{e}clet number, is the ratio of a diffusion timescale, $\delta^2/D$, to the advection timescale, $L/(2q/\delta)$. Note that $Pe=\frac{\delta}{L}\left(\frac{q}{\delta}\frac{\delta}{D}\right)=\frac{\delta}{L}Pe_c$, where $Pe_c=\frac{q}{D}\backsim u\frac{\delta}{D}$ is the conventional definition of the P\'{e}clet number. Furthermore, $\Lambda$ and $Pe$ depend only on particle and flow properties, which are either given parameters (e.g.\ $L$ and $q$) or can be directly measured from experiments and simulations (e.g.\ $\delta$, $S$ and $D$). %[Let's try to get some empirical equations]

\subsection{Boundary conditions}
\label{bc}

As mentioned in \S \ref{intro}, we restrict our attention to the steady filling stage which occurs when the heap extends  to the downstream bounding endwall and rises with uniform velocity. At the inlet boundary ($\tilde{x}=0$), $c_{s}(0,\tilde z)=c_{l}(0,\tilde z)=0.5$ for initially well-mixed particles. At the top and bottom boundaries of the flowing layer ($\tilde{z}=-1$ and 0), the segregation flux equals the diffusive flux, as in \citet{Gray2006},
\begin{equation}\label{eq:TopBotBC}
\Lambda(1-\tilde{x})\left[g(-1)c_i(1-c_i)\right]=\frac{1}{{Pe}}\frac{\partial c_i}{\partial \tilde{z}},
\end{equation}
which indicates that particles exit the heap through the bottom of the flowing layer at $w=-v_r{\rm cos}\alpha$ only through advection due to the mean flow  (in the moving reference frame). At the downstream boundary ($\tilde{x}=1$), flow is parallel to the wall ($\tilde{u}(1,\tilde{z})=0$) and, since both diffusion and segregation act only in the $z-$direction, no boundary condition is required.

The flux boundary condition at $z=-\delta$ (equation~\eqref{eq:TopBotBC}) is necessary since the streamwise velocity and the shear rate are small, but nonzero below $z=-\delta$ (see figure~\ref{kinematics}(c)). For $z<-\delta$, we assume that the streamwise velocity, the percolation velocity, and diffusion are negligible. Consequently, the concentrations directly below $z=-\delta$ are the same as those at $z=-\delta$. For mass to be conserved in the heap
\begin{equation}
\frac{\intop_{0}^{L}w(x,-\delta)c_{i}(x,-\delta)dx}{\intop_{0}^{L}w(x,-\delta)dx}=0.5,
\end{equation}
when equal concentrations of small and large particles enter the flowing layer from the feed zone. For this condition to be met, the diffusion and the segregation fluxes must be equal at $z=-\delta$, so that particles only exit the domain via advection.

Moreover, if the shear rate at the bottom of the flowing layer, represented by $g(-1)$ (equation~\eqref{eq:TopBotBC}), is small, and, if $Pe$ is not too large, $\partial c/\partial \tilde{z}\approx0$ at $z=-\delta$, consistent with the assumption that particle concentrations are effectively constant below $z=-\delta$. If $g(-1)$ is not small at the bottom of the flowing layer, then $\partial c/\partial \tilde{z}$ could be large there, which is non-physical, because particles move slowly near the bottom of the flowing layer. A rapid change in particle concentrations at the bottom of the flowing layer can occur when $Pe$ is large (equation~\eqref{eq:TopBotBC}). As $Pe\rightarrow\infty$, a discontinuity (or shock) in the concentration of small/large particles at the bottom of the flowing layer occurs \citep{Gray2009}. In \S \ref{comparison}, we examine the differences between the exponential velocity profile (equation~\eqref{w_z_exp}), where $g(-1)$ is small ($g(-1)=g(0)/10$), and the linear velocity profile (equation~\eqref{w_z_linear}), where $g(-1)$ is not small ($g(-1)=g(0)$). In the end, an exponential velocity profile with boundary condition \eqref{eq:TopBotBC} matches DEM simulations and experiments quite well.

\subsection{Numerical Method}
\label{sec:numeric_method}

Equation~\eqref{eq:ND} is solved for the steady state using an operator splitting method, which divides the computation into an advection step and a combined segregation and diffusion step, each of which is easier to solve than the full problem. Operator splitting schemes for advection-diffusion equations have been used previously to study the diffusion of a magnetic field in fast dynamos \citep{OttOpSp}, tracer trajectories in turbulent flows \citep{Jones}, and strange eigenmodes in granular flows \citep{Christov11}. Recently, a study by \citet{Schlick13} verified the accuracy of operator splitting techniques in advection-diffusion problems.

Similar to the approach in \citet{Schlick13}, to evolve the system from time $\tilde{t}=m\Delta \tilde{t}$ to $t=(m+1)\Delta \tilde{t}$, we first solve the advection step
\begin{equation}
\frac{\partial c^{*}}{\partial\tilde{t}}=-\tilde{u}\frac{\partial c^{*}}{\partial\tilde{x}}
-\tilde{w}\frac{\partial c^{*}}{\partial\tilde{z}}\qquad \tilde{t}\in[m\Delta \tilde{t},(m+1)\Delta \tilde{t}],
\label{eq:Adv}
\end{equation}
for $c^{*}((m+1)\Delta \tilde{t})$ using $c(m\Delta \tilde{t})$ as the initial condition. Next, the segregation and diffusion step is solved,
\begin{equation}
\frac{\partial c}{\partial\tilde{t}}=-\Lambda(1-\tilde{x})\frac{\partial}{\partial\tilde{z}}\left[g(\tilde{z})c(1-c)\right]
+\frac{\partial}{\partial\tilde{z}}\left(\frac{1}{Pe}\frac{\partial c}{\partial \tilde{z}}\right),\qquad \tilde{t}\in[m\Delta \tilde{t},(m+1)\Delta \tilde{t}],
\label{eq:SegDiff}
\end{equation}
using $c^{*}((m+1)\Delta \tilde{t})$ as the initial condition. To solve each step, the domain is subdivided into an $N_{x}$ by $N_{z}$ grid. As in \citet{Christov11} and \citet{Schlick13}, equation~\eqref{eq:Adv} is solved with a matrix mapping method. The matrix mapping method uses an $N_{x}N_{z}\times N_{x}N_{z}$ matrix, $\mathbf{\Phi}_{\Delta \tilde{t}}$, where each entry $\mathbf{\Phi}_{\Delta \tilde{t}}^{(a,b)}$ represents the proportion of material in cell $a$ carried by the velocity field $\bm{u}$ from cell $b$ in time $\Delta \tilde{t}$. Therefore, if $\bm{c}$ is a $N_{x}N_{z}\times1$ column vector of the concentrations in each grid cell, then $\bm{c}(\tilde{t}_{0}+\Delta \tilde{t})=\mathbf{\Phi}_{\Delta \tilde{t}}\bm{c}(\tilde{t}_{0})$. For more details on matrix mapping methods, see \citet{Anderson09,Anderson09_eig} and \citet{Schlick13}.

The segregation and diffusion step (equation~\eqref{eq:SegDiff}) is solved using the implicit Crank--Nicolson method implemented on each column in the $N_{x}\times N_{z}$ grid (constant $\tilde{x}$, $-1<\tilde{z}<0$). Since the segregation term contains a nonlinearity, the method of successive approximations (or inner iterations) is used as in \citet{Ames}.

The utility of this scheme is that each column in the $N_{x}\times N_{z}$ grid depends only on the columns to its left (smaller $\tilde{x}$) in steady state. This is because the matrix mapping method used to solve the advection step depends solely on the concentration profile upstream, which in this case is only columns to the left  of the given column (as $\tilde{u}\ge0$). Therefore, to solve equation~\eqref{eq:ND} in steady state, the concentration in each column is determined sequentially, starting with the second column (the first column's concentration is determined by the inlet condition).

\section{Model predictions}
\label{results}

\subsection{Comparison with experiments and simulations}
\label{comparison}
\begin{figure}
 \centerline{\includegraphics[scale=.7]{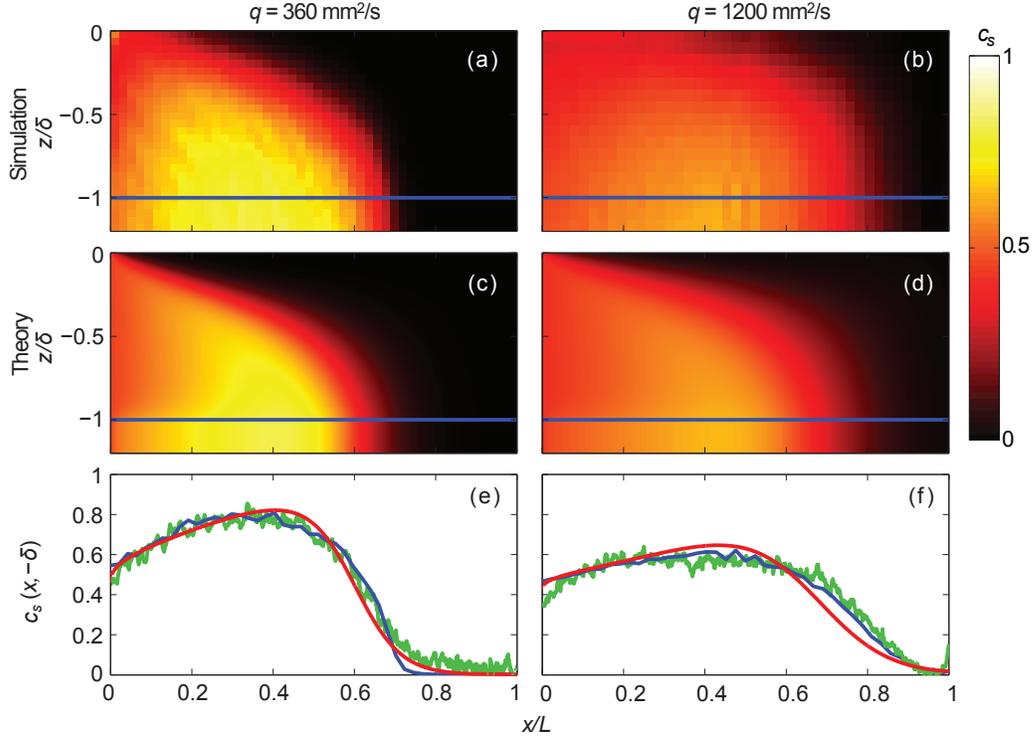}}
\caption{Small particle concentration, $c_s$, from (a,b) simulation and (c,d) theory. (e, f) $c_s$ at the bottom of the flowing layer as a function of streamwise position, $x/L$, calculated from simulation (black, blue online), theory (dark grey, red online), and experiment (light grey, green online). For the lower flow rate (left column), $S=0.19$~mm, $L=490$~mm, $\delta=11$~mm, $q=360~\text{mm}^{2}/$s, $D=0.8~\text{mm}^{2}/$s, $\Lambda=0.78$, and $Pe=19$. For the higher flow rate (right column), $S=0.18$~mm, $L=430$~mm, $\delta=14$~mm, $q=1200~\text{mm}^{2}/$s, $D=2.83~\text{mm}^{2}/$s, $\Lambda=0.4$, and $Pe=28$.}
\label{fig:Compare_sim_theory}
\end{figure}
To validate the model, we compare steady state solutions of equation~\eqref{eq:ND} with DEM simulation results and our previous experimental results \citep{Fan2012} at the same operating conditions (feed rate, size ratio, system size, and inlet condition). In DEM simulations and experiments, $c_{s}(0,\tilde{z})<0.5$, since more small particles than large particles fall out of the flowing layer in the feed zone ($\tilde{x}<0$). Therefore, the inlet condition in theory, $c_{0}=c_{s}(0,\tilde{z})$, is calibrated such that the flux of particles into the domain at the upstream end of the flowing layer is the same for both theory and simulation. Figure~\ref{fig:Compare_sim_theory} shows comparisons for a mixture of 1~mm and 2~mm diameter particles at two different feed rates: $q=360~\text{mm}^2/\text{s}$ (left column) and $q=1200~\text{mm}^2/\text{s}$ (right column). For the theoretical predictions, the values of the two dimensionless parameters $\Lambda$ and $Pe$ are calculated based on operating conditions ($q$ and $L$) and direct measurements from DEM simulations ($S$ and $D$). The thickness of the flowing layer, $\delta$, is determined based on the profiles of the streamwise velocity in the depth direction \citep{Fan2013}. The results in figure~\ref{fig:Compare_sim_theory} are based on the exponential streamwise velocity profiles (equation~\eqref{w_z_exp}) and a constant diffusion coefficient.

Figures~\ref{fig:Compare_sim_theory}(a)-(d) show the volume concentration contours of small particles at two feed rates. DEM simulations and theoretical predictions agree quite well in both cases. Segregation occurs in the flowing layer ($-1 \le z/\delta \le 0$, above the solid line in each sub-figure). Large particles segregate toward the free surface, are advected to the end of the flowing layer, and deposit onto the static bed in the downstream region (black region). Small particles percolate toward the bottom of the flowing layer and deposit onto the static bed in the upstream region (light orange region). In the creeping region ($-1.2\le z/\delta<-1$), particle concentrations are nearly invariant in the normal direction. At the higher feed rate, the degree of segregation decreases in that fewer large particles segregate to the downstream region (figures~\ref{fig:Compare_sim_theory}(b, d)) compared to the lower feed rate case (figures~\ref{fig:Compare_sim_theory}(a, c)).

Figures~\ref{fig:Compare_sim_theory}(e, f)  further compare small particle streamwise concentration profiles between theory, simulation and experiment \citep{Fan2012} at the bottom of the flowing layer ($\tilde z=-1$). At both feed rates, the theoretical predictions match well both experiment and simulation.

\begin{figure}
 \centerline{\includegraphics[scale=.6]{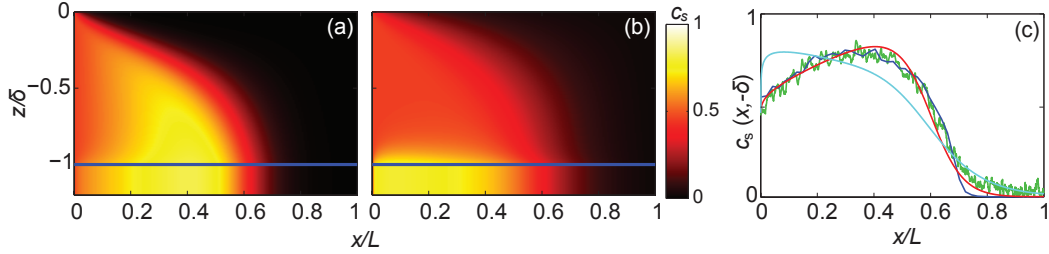}}
 \caption{Effect of different streamwise velocity profiles on theoretical prediction. (a,b) Small particle concentration, $c_s$, fields for (a) exponential and (b) linear velocity profiles with 1 and 2 mm diameter particles at $Q=4.57\times 10^3$ ~mm$^3$/s. (c) Comparison of the profiles of $c_s$ at the bottom of the flowing layer in the $x-$direction for the theoretical predictions [exponential velocity profile: dark grey (red online), linear velocity profile: lightest grey (cyan online)], experiment [lighter grey (green online)], and simulation [black (blue online)] for $\Lambda=0.78$ and $Pe$ = 19.}
\label{fig_sensitivity}
\end{figure}

To examine how the mean flow velocity profile affects the accuracy of the model predictions, we compare solutions to equation~\eqref{eq:ND} for both the exponential model (equation~\eqref{w_z_exp}) and the linear model (equation~\eqref{w_z_linear}) using the same values for $\Lambda$ and $Pe$ in each case. The key difference between the two velocity profiles is that the percolation velocity is constant from the free surface to the bottom of the flowing layer at each streamwise location for the linear streamwise velocity profile, while it decreases exponentially for the exponential streamwise velocity profile. This can significantly affect particle distributions at the same operating conditions, as shown in figure~\ref{fig_sensitivity}. As predicted in \S \ref{bc}, the linear velocity profile produces a rapid change in particle concentrations at the bottom of the flowing layer, resulting in an anomalous layer of small particles just above $z=-\delta$ (figure~\ref{fig_sensitivity}(b)), while the exponential velocity profile does not (figure~\ref{fig_sensitivity}(a)). As discussed in \S \ref{bc}, the rapid change in particle concentration at the bottom of the flowing layer for the linear velocity profile is due to the large shear rate at $z=-\delta$, while for the exponential velocity profile, the shear rate decreases with depth in the flowing layer consistent with the velocity profile in figure~\ref{kinematics}(a). Hence, the linear velocity profile produces non-physical results. In addition, theoretical predictions of small particle concentration using an exponential velocity profile with a constant percolation velocity also do not match experiments and simulations.

The theoretical prediction for the small particle concentration at the bottom of the flowing layer based on the exponential velocity profile matches both simulation and experiment, as seen in figure~\ref{fig_sensitivity}(c). In contrast, $c_s$ from the linear velocity profile is over-predicted in the upstream region due to an overestimate of the segregation fluxes near $z=-\delta$. Note that \citet{Wiederseiner2011} found that in inclined chute flow, when all three driving factors -- advection, segregation, and diffusion -- are accounted for in the transport equation, only qualitative agreement is obtained between the theoretical prediction and experiment in the upstream region of the flow. They speculated that the cause was an inaccurate description of the complicated streamwise velocity profiles in the upstream region, which is supported by the results in figure~\ref{fig_sensitivity}.

When the diffusion coefficient is allowed to vary spatially with the shear rate, only marginally improved predictions of experiment and simulation are achieved (see Appendix \ref{sensitivity}).
Therefore, we use the mean velocity field based on an exponential velocity profile (equation~\eqref{w_z_exp}) and a constant $D$ throughout this paper.

\subsection{Influence of $\Lambda$ and $Pe$ on particle configuration}

\begin{figure}
 \centerline{\includegraphics[scale=.7]{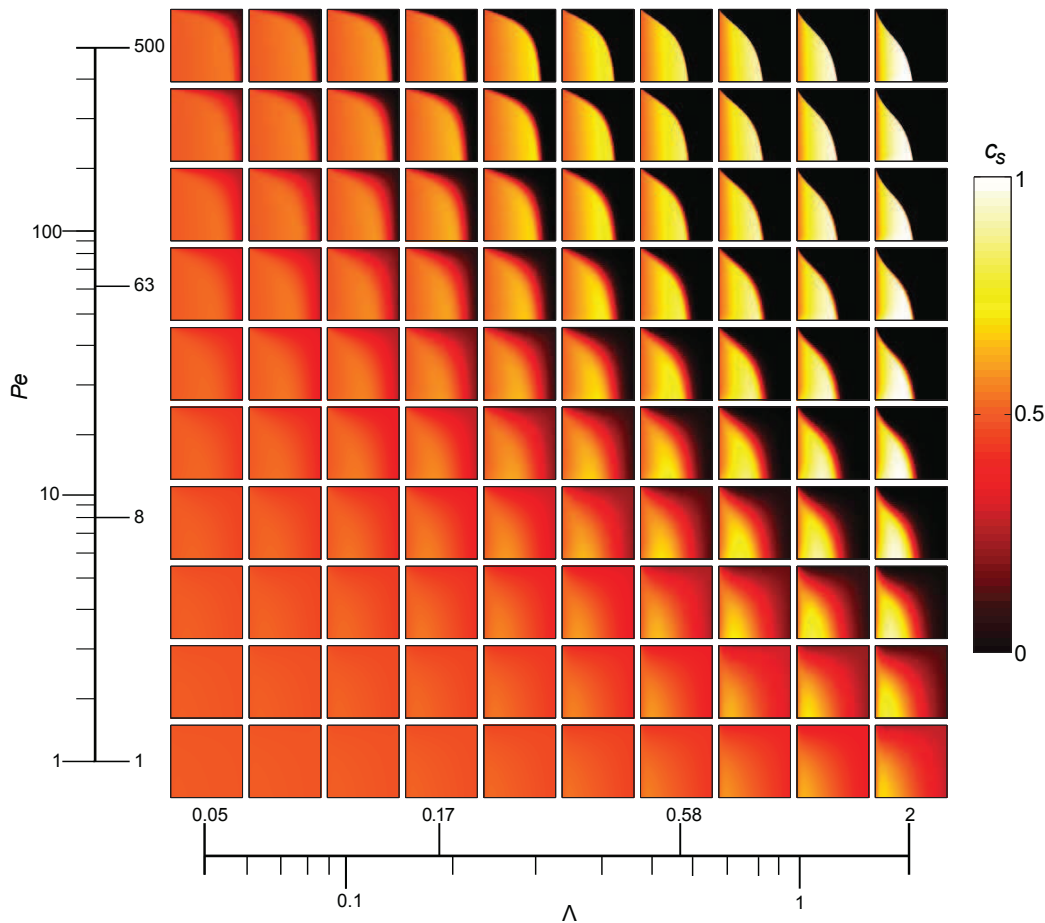}}
\caption{Small particle concentration contours in the flowing layer for different $\Lambda$ and $Pe$ for the well-mixed inlet condition, showing transition from mixed (bottom left) to segregated states (top right). Each box shows the entire flowing layer domain ($0 \le \tilde x \le 1$ and $-1 \le \tilde z \le 0$).}
\label{fig:AllStates}
\end{figure}

We now systematically investigate the effect of $\Lambda$ and $Pe$ on the particle configuration based on the theoretical model. In figure \ref{fig:AllStates}, an array of contour maps of small particle concentration in the flowing layer (like those in figures \ref{fig:Compare_sim_theory}(a)-(d)) are shown for a wide range of $\Lambda$ and $Pe$. A strongly segregated state occurs at high $\Lambda$ and high $Pe$ (top right), and a well-mixed state occurs at low $\Lambda$ and low $Pe$ (bottom left). The transition from segregated states to mixed states can be achieved by decreasing either $\Lambda$ or $Pe$. This corresponds to decreasing percolation (by decreasing $\Lambda$) or increasing diffusion (by decreasing $Pe$) to obtain greater mixing. However, there is a subtle but \textit{non-trivial} difference between these two scenarios. At high $Pe$ and low but non-zero $\Lambda$ (left top), there is a small region of mostly large particles at the end of flowing layer, a well-mixed region upstream, and a sharp transition between the two regions. This exactly matches our previous experiments \citep{Fan2012} at high feed rates. However, at low $Pe$ and high $\Lambda$ (right bottom), the small particle concentration decreases gradually along the streamwise direction and the region of pure large particles does not occur, which has also been observed in previous experiments \citep{Goyal2006,Fan2012}. This difference can be attributed to the advection effect, which will be discussed later in this section and in \S \ref{discussion}.

\begin{figure}
 \centerline{\includegraphics[scale=0.8]{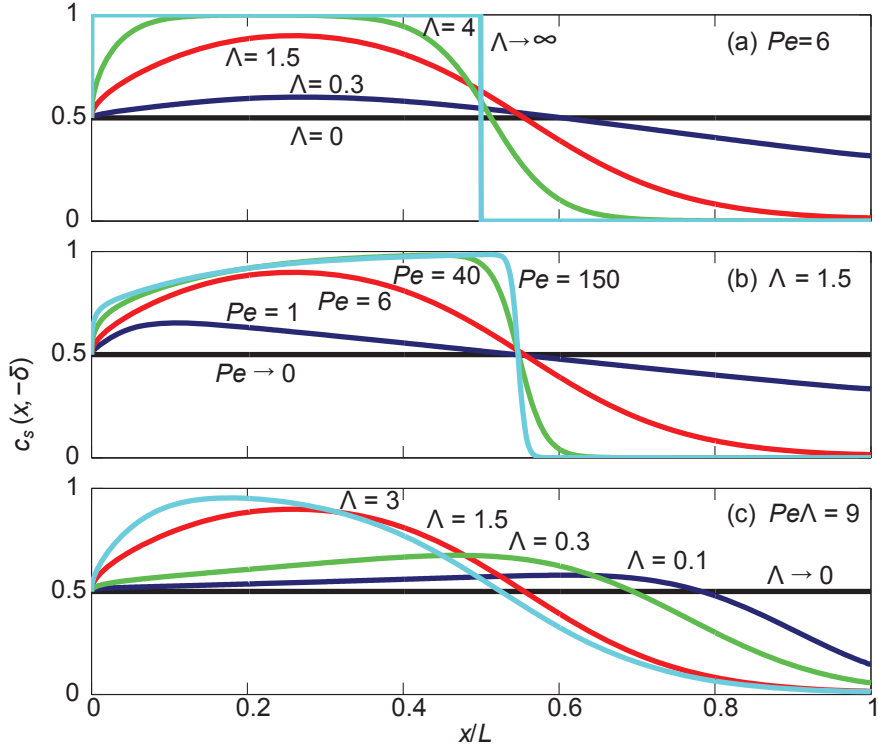}}
\caption{Small particle concentration at the bottom of the flowing layer as a function of streamwise location $x/L$ for (a) $Pe=6$, (b) $\Lambda=1.5$, and (c) Pe$\Lambda=9$ based on the steady state solution of equation~\eqref{eq:ND}.}
\label{fig:Cs_Bottom_FL}
\end{figure}

To further investigate the effects of $\Lambda$ and $Pe$, we consider the small particle concentration profiles at the bottom of the flowing layer in the streamwise direction (as in figures \ref{fig:Compare_sim_theory}(e, f)) for different combinations of $\Lambda$ and $Pe$. Figure~\ref{fig:Cs_Bottom_FL}(a) shows the effects of changing $\Lambda$ for constant $Pe$ ($Pe$ = 6). When $\Lambda\rightarrow \infty$, segregation dominates both diffusion and advection, so that small and large particles completely segregate immediately after entering the flowing layer. This results in a completely segregated pattern, where all small particles accumulate in the upper half of the heap ($x/L<0.5$) and all large particles are advected to the lower half of the heap ($x/L>0.5$). In contrast, when $\Lambda=0$, no segregation occurs and advection and diffusion effects keep the entire flowing layer mixed (similar to the left bottom corner of figure~\ref{fig:AllStates}). Between these two limits ($0<\Lambda<\infty$), the concentration of small particles increases moving downstream to a maximum value in the upstream portion of the flowing layer and then gradually decreases. For large enough $\Lambda$, the small particle concentration eventually decreases to 0 in the downstream portion of the flowing layer, leaving only large particles at the end of the flowing layer, as reported by \cite{Fan2012}.

Figure~\ref{fig:Cs_Bottom_FL}(b) shows the effect of changing $Pe$ for constant $\Lambda$ ($\Lambda=1.5$). When $Pe\rightarrow 0$ ($D \rightarrow \infty$), diffusion dominates segregation and advection, producing a perfectly mixed state in the entire flowing layer. In contrast, for $Pe>150$, the diffusion effect becomes weaker, and the particle concentration profile is similar to that at high values of $\Lambda$ in figure~\ref{fig:Cs_Bottom_FL}(a), where the two species segregate nearly completely. For intermediate values of $Pe$ ($0<Pe<150$), the small particle concentration increases moving downstream until it reaches a maximum value and then gradually decreases, similar to figure~\ref{fig:Cs_Bottom_FL}(a) for moderate values of $\Lambda$. However, the location of the maximum value of the small particle concentration moves downstream as $Pe$ increases.

To better demonstrate the advection effect on particle configuration, we vary both $\Lambda$ and $Pe$ while keeping their product constant. Constant $\Lambda Pe$ indicates that the ratio between the segregation and diffusion effects remains the same. When $Pe$ (or, alternatively $\Lambda$) changes, the advection effect will change correspondingly. Figure~\ref{fig:Cs_Bottom_FL}(c) indicates that when $Pe$ increases and $\Lambda$ decreases (corresponding to moving from the bottom right region to the upper left region of figure~\ref{fig:AllStates}), the advection effect becomes stronger, so a better mixed state is achieved and the location of the maximum small particle concentration moves further downstream. This occurs because strong advection preserves the upstream particle distribution. In other words, the particles remain mixed so small particles are advected farther down the heap. Alternatively, if the mixture is unmixed at the flow inlet ($c_{s}(0,z)\ne c_{l}(0,z)$), strong advection can preserve the unmixed state, a case which is discussed in detail in the next section.

\section{Interplay of segregation, advection, and diffusion}
\label{discussion}

%\subsection{}
%\label{competition}

The nondimensional governing equation~\eqref{eq:ND} indicates that the two dimensionless parameters, $\Lambda$ and $Pe$, control particle configuration through the interplay of advection, segregation, and diffusion in bounded heap flow. Segregation, controlled only by $\Lambda$, separates small and large particles in the normal direction. Diffusion, controlled only by $Pe$, mixes small and large particles across concentration gradients and hinders segregation. Advection, however, is manifested in both $\Lambda$ and $Pe$ (e.g.\ the advection effect is strong if $\Lambda$ is small and $Pe$ is large). Strong advection tends to maintain the particles in the same mixture conditions as at the inlet. Here, we examine the influence of the three mechanisms by considering their time scales.
%\footnote{Since this is a Graetz problem, characteristic spatial scales could be used as readily as time scales. We prefer to use time scales, as they are more easily interpreted.}.

\begin{figure}
 \centerline{\includegraphics[scale=.75]{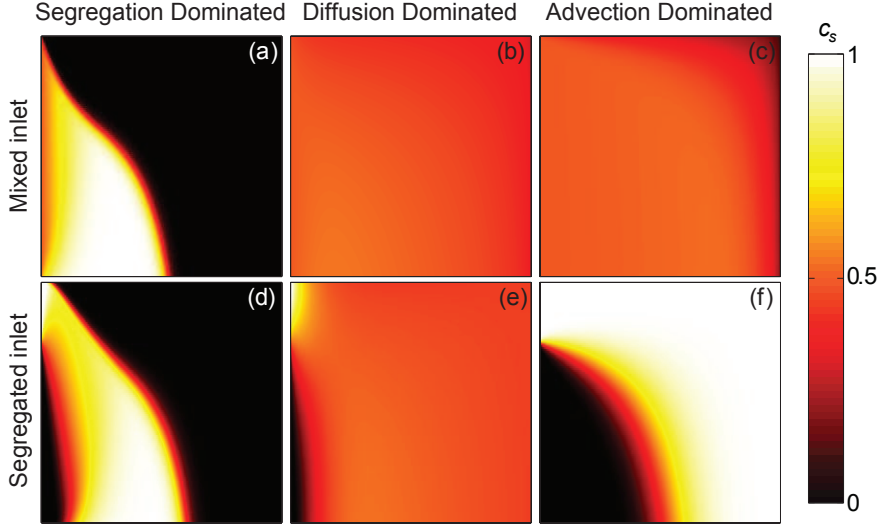}}
\caption{Contours of small particle concentration in the flowing layer for (a-c) a mixed inlet condition ($c_{s}(0,\tilde{z})=c_{l}(0,\tilde{z})=0.5$) and (d-f) a segregated inlet condition (equation~\eqref{eq:ReverseBC}). (a, d) Segregation-dominated ($\Lambda=2.5$, $Pe=20$, $\tilde{t}_{a}=2$, $\tilde{t}_{d}=2$, and $\tilde{t}_{s}=0.2$). (b, e) Diffusion-dominated ($\Lambda=0.25$, $Pe=2$, $\tilde{t}_{a}=2$, $\tilde{t}_{d}=.2$, and $\tilde{t}_{s}=2$). (c, f) Advection-dominated ($\Lambda=0.025$, $Pe=200$, $\tilde{t}_{a}=2$, $\tilde{t}_{d}=20$, and $\tilde{t}_{s}=20$).}
\label{fig:AdvSegDiff_dominate}
\end{figure}

The segregation timescale, $t_{s}$, is proportional to $\delta/w_{p}$, where $w_{p}$ is the percolation velocity from equation~\eqref{percolation_v}. Nondimensionalizing $t_{s}$ using equation~\eqref{normalization} (i.e. $\tilde{t}_{s}=t_{s}/(\delta L/2q)$) yields $\tilde{t}_{s}\sim1/\Lambda$. Similarly, the diffusion timescale, $t_{d}\sim \delta^{2}/D$, takes the dimensionless form $\tilde t_{d}\sim Pe$. The advection timescale, $t_{a}\sim L/u$ (or, alternatively, $t_{a}\sim \delta/w$), is nondimensionalized to $\tilde{t}_{a}\sim 1$, since equation~\eqref{normalization} defines $t_{a}\sim\delta L/2q$ as the advection timescale. The order of magnitude for these dimensionless timescales can be estimated (see Appendix~\ref{time_scales}) as:

\begin{equation}
\tilde{t}_{s}=0.5/\Lambda,\quad\quad\tilde{t}_{d}={Pe}/10,\quad {\rm and}\quad \tilde{t}_{a}=2.
\label{eq:timescales}
\end{equation}

The effects of the three mechanisms on particle distributions can be elucidated by controlling the above timescales for two different flow inlet conditions as shown in figure~\ref{fig:AdvSegDiff_dominate}. In addition to the well-mixed inlet condition ($c_{s}(0,\tilde{z})=c_{l}(0,\tilde{z})=0.5$), we also consider an ``inverted'' segregated inlet condition, where small particles are above large particles at the flow inlet:
\begin{equation}
c_{s}(0,\tilde{z})=1-c_{l}(0,\tilde{z})=\begin{cases}
1, & -0.25 \le \tilde{z} \le 0\\
0, & -1\le \tilde{z}<-0.25.
\end{cases}
\label{eq:ReverseBC}
\end{equation}
For this inlet condition, the fluxes of small and large particles entering the flowing layer at $\tilde{x}=0$ are approximately equal.

Figures~\ref{fig:AdvSegDiff_dominate}(a, d) show that when the segregation effect dominates ($\tilde t_{s}$ one order of magnitude smaller than $\tilde t_{d}$ and $\tilde t_{a}$), small and large particles segregate almost completely for both inlet conditions, except for some large particles that initially deposit into the static bed at small $x$ for the segregated inlet condition. In other words, even though the particles in figure~\ref{fig:AdvSegDiff_dominate}(d) begin with the small particles above the large ones, segregation is so strong that the small particles still deposit on the heap upstream of the large ones. When diffusion dominates ($\tilde t_{d}$ one order of magnitude smaller than $\tilde t_{s}$ and $\tilde t_{a}$), particles are mixed in most of the flowing layer for both inlet conditions (figures~\ref{fig:AdvSegDiff_dominate}(b, e)). However, as figures~\ref{fig:AdvSegDiff_dominate}(c, f) show, when advection dominates ($\tilde t_{a}$ one order of magnitude smaller than $\tilde t_{s}$ and $\tilde t_{d}$), particle configurations are quite different between the two inlet conditions: a well-mixed inlet condition produces a well mixed state and a segregated inlet condition produces a segregated state (inverted from those in figures~\ref{fig:AdvSegDiff_dominate}(a, d)). These results demonstrate that strong advection preserves the inlet condition in bounded heap flow, as particles have little time to segregate or diffuse before leaving the flowing layer.

\begin{figure}
 \centerline{\includegraphics[scale=0.7]{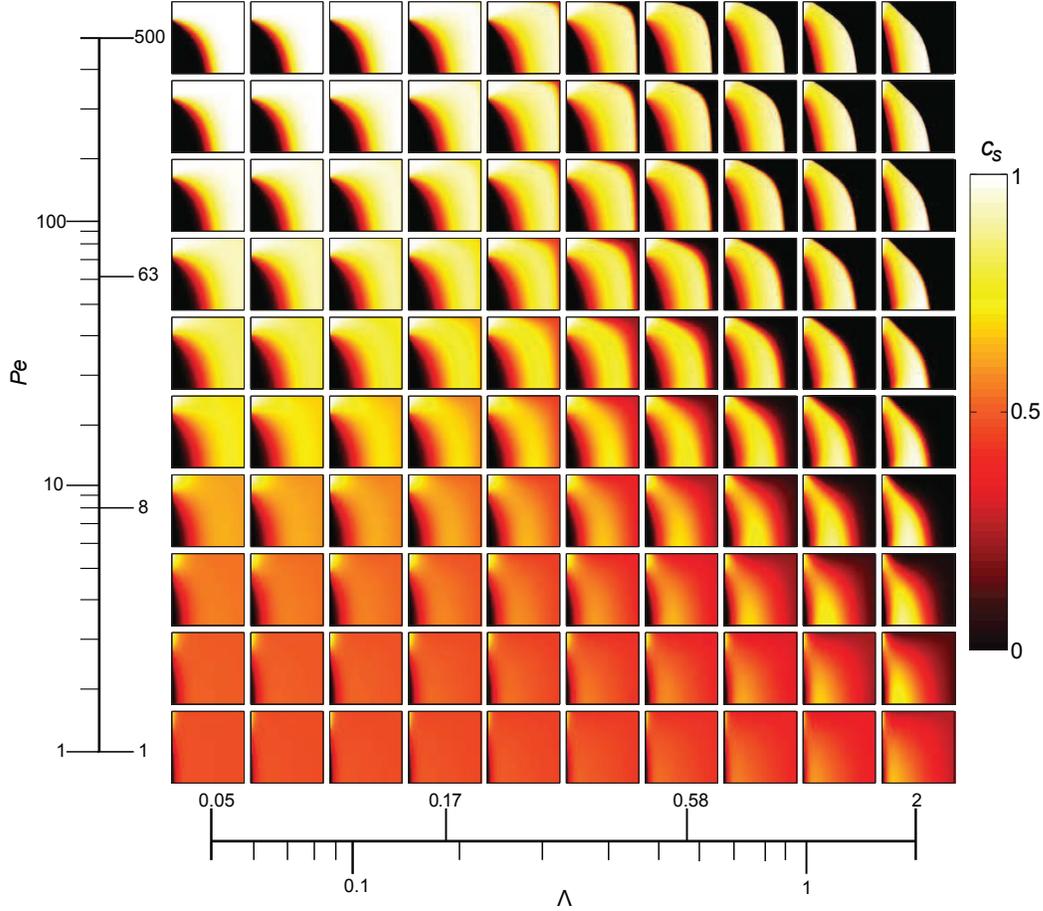}}
\caption{Small particle concentration contours in the flowing layer for different $\Lambda$ and $Pe$ for the segregated inlet condition (equation \eqref{eq:ReverseBC}). Each box shows the entire flowing layer domain ($0 \le \tilde x \le 1$ and $-1 \le \tilde z \le 0$).}
\label{figure:SBC}
\end{figure}

To further illustrate the interplay of advection, segregation, and diffusion, figure \ref{figure:SBC} shows the $\Lambda-Pe$ space map (similar to figure~\ref{fig:AllStates}) for the segregated inlet condition. The initially segregated state persists in the advection-dominated regime (high $Pe$ and low $\Lambda$) and a well-mixed state is obtained in the diffusion-dominated regime (low $Pe$ and low $\Lambda$), in contrast to the well-mixed states in both regimes for the mixed inlet condition (see figure~\ref{fig:AllStates}). When $\Lambda$ increases, percolation dominates so that a final segregated state opposite to the segregated inlet boundary condition occurs, except at very low $Pe$.

\begin{figure}
 \centerline{\includegraphics[scale=0.85]{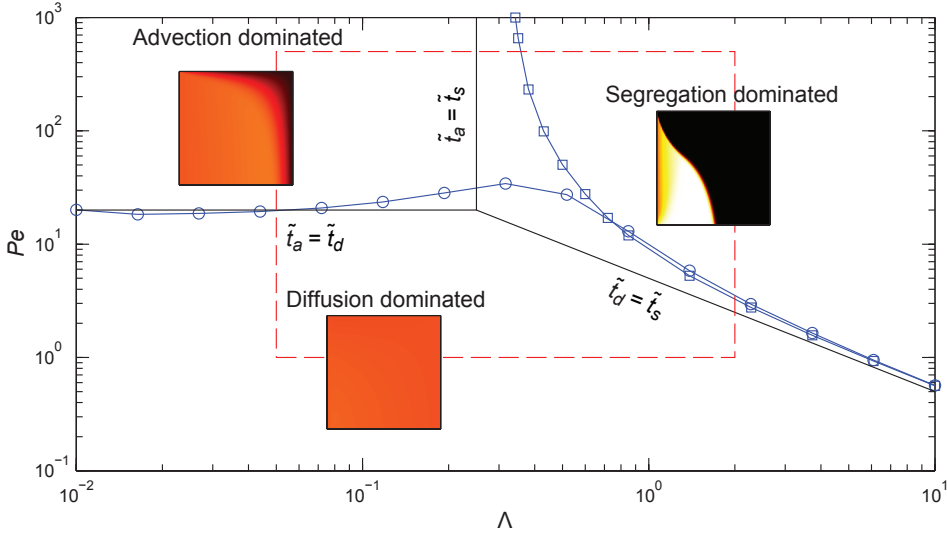}}
\caption{$\Lambda-Pe$ parameter space illustrating different segregation regimes. Black lines indicate where pairs of time scales (from equation \eqref{eq:timescales}) are equal and divide parameter space into regions dominated by either advection, diffusion, or segregation. In each region, the representative concentration profile from figure \ref{fig:AdvSegDiff_dominate} for the mixed inlet condition is shown. Blue curves are contours of $I_{d}=1/e$ for mixed (\small$\square$) and segregated ({\large$\circ$}) inlet conditions. Red (dashed) box indicates portion of parameter space shown in figures~\ref{fig:AllStates} and \ref{figure:SBC}.}
\label{fig:Lambda_Pecrit}
\end{figure}

Using the timescales of advection, segregation, and diffusion, it is possible to investigate how each mechanism affects particle configurations in bounded heap flow in $\Lambda-Pe$ space (see figure~\ref{fig:Lambda_Pecrit}). The space can be divided into three regimes in which one mechanism dominates. The boundaries between these regimes (black lines in figure~\ref{fig:Lambda_Pecrit}) are determined by equating pairs of timescales, i.e.\ $\tilde t_{a}=\tilde t_{s}$, $\tilde t_{d}=\tilde t_{a}$, and $\tilde t_{s}=\tilde t_{d}$, using the values in equation~\eqref{eq:timescales}. The goal is to determine whether boundaries between different regimes match the transition between different particle configurations shown in figures~\ref{fig:AllStates} and \ref{figure:SBC}.

To quantify the global mixing at steady state, we use the Danckwerts intensity of segregation~\citep{Id}, defined as
\begin{equation}
I_{d}=\frac{1}{L\bar{c}(1-\bar{c})}\intop_{0}^{L}[c(x,-\delta)-\bar{c}]^{2}dx.
\end{equation}
Here, $I_{d}$ measures the amount of mixing at the bottom of the flowing layer (i.e.\ the particles that deposit onto the static heap) and $\bar c=\frac{1}{L}\intop_{0}^{L}c(x,-\delta)dx=0.5$ is the mean particle concentration at the bottom of the flowing layer. By definition, $c_s+c_l=1$, so that $I_d$ is the same for both small and large particles. For a completely segregated final state, $I_{d}=1$, and for a completely mixed final state, $I_{d}=0$. In our previous experiments~\citep{Fan2012}, we quantified the degree of segregation using $\Delta L/L$, where $\Delta L$ is the length of the flowing layer at the downstream end occupied by large particles. While convenient in experiments, this metric does not adequately capture mixing in the heap in several cases. For a relatively well-mixed final state with no distinct band of large particles at the end of the heap (e.g.\ bottom left of figure~\ref{fig:AllStates}), $\Delta L/L=0$ and fails to distinguish subtle differences in concentration profiles. Furthermore, if $\Lambda$ is small and $Pe$ is large (e.g.\ top left of figure~\ref{fig:AllStates}), a narrow band of large particles at the end of the heap makes $\Delta L/L>0$, even though the heap is well-mixed everywhere else. In comparison, the Danckwerts measure is appropriate in all cases.

In figure~\ref{fig:Lambda_Pecrit}, the curve of constant $I_d=1/e$ for a well-mixed inlet condition ($\square$) divides the phase diagram into two parts: a segregated state in the right top portion with higher $I_d$ and a mixed state with lower $I_d$ elsewhere, consistent with the results in figure~\ref{fig:AllStates} (corresponding to the red dashed box in figure~\ref{fig:Lambda_Pecrit}). The boundary between segregation-dominated and diffusion-dominated regimes based on the time scales ($\Lambda Pe$=5) nearly overlays the curve of constant $I_d$ for large values of $\Lambda$. The boundary between segregation-dominated and advection-dominated regimes ($\Lambda$=0.25) also qualitatively matches the curves of constant $I_d$ for large values of $Pe$. Note that the cutoff value $1/e$ for $I_d$ is arbitrary. Other cutoff values (e.g.\ 0.2, 0.3, and 0.4) do not substantially influence the results in figure~\ref{fig:Lambda_Pecrit}.

For the well-mixed inlet condition, both strong advection and diffusion effects lead to a well-mixed heap, so the boundary between these two effects cannot be distinguished using $I_d$. However, the transition between the advection-dominated regime and diffusion-dominated regime can be identified by plotting the curve of constant $I_d=1/e$ ($\circ$ in figure~\ref{fig:Lambda_Pecrit}) for the segregated inlet condition. This curve represents the boundary between the advection-dominated regime and the diffusion-dominated regime for small $\Lambda$ ($<0.25$), and the boundary between the diffusion-dominated regime and the segregation-dominated regime for large $\Lambda$ ($>0.25$). This curve again qualitatively matches the boundaries based on the time scales. Moreover, for high $\Lambda$ ($>1$), the curves of constant $I_{d}$ for the segregated inlet condition and mixed inlet condition approach each other as $\Lambda$ increases, because the advection effect becomes weaker and the inlet condition cannot persist for long.

\section{Conclusion}
\label{conclusion}

In this paper we have developed a predictive model for the spatial distribution of bidisperse granular materials in bounded heap flow using a classical transport formalism. The theoretical predictions match well with experimental and simulation results. The model includes the effects of three different mechanisms -- advection due to mean flow, segregation due to percolation, and diffusion due to random particle collisions. Compared with previous predictive models \citep{Shinohara1972,Boutreux1996}, the model presented here is based on an understanding of the kinematics of bounded heap flow and has no arbitrarily adjustable fitting parameters. Instead, particle configurations are controlled by two dimensionless parameters: $\Lambda=SL/\delta^2$ and $Pe=2q\delta/DL$. Both parameters are functions of physical control parameters (e.g.\ feed rate, $q$, and flowing layer length, $L$) and kinematic parameters that can be measured from experiments or simulations (e.g.\ diffusion coefficient, $D$, percolation length scale, $S$, and flowing layer thickness, $\delta$). Particle configurations can be controlled by $\Lambda$ and/or $Pe$ through the physical control parameters such as $S$ (by changing size ratio), $L$, or $q$. Furthermore, these two dimensionless parameters reveal the physical mechanisms observed in previous experiments \citep{Fan2012}. $\Lambda$ describes the interplay between segregation and advection (essentially the same as the dimensionless time scale $\tilde t$ in \citet{Fan2012}), and $Pe$ represents the interplay between advection and diffusion. A parametric study of $\Lambda$ and $Pe$ and a dimensional analysis of the timescale of the three different driving mechanisms show how particle configurations in bounded heap flow depend on the interplay of advection, segregation, and diffusion.

The kinematic parameters ($D$, $S$, and $\delta$) can be measured from simulations and experiments, but their relationship with the physical control parameters ($q$, $L$, and the particle sizes, $d_s$ and $d_l$) is not yet clear. We are currently investigating whether and, if so, how $\Lambda$ and $Pe$ can be determined solely from the physical control parameters.

The theoretical framework for modeling segregation and mixing of granular flows described here is not limited to quasi-2D bounded heap flow, but can be adapted for other flow geometries (including three-dimensional systems) as long as the flow kinematics are accurately determined. This is particularly useful for flows with complicated kinematics such as rotating tumbler flow, where rich segregation-driven patterns have been observed \citep{Ottino2000,Meier2007}. New challenges arise, though. In a thin rotating cylindrical tumbler, there are gradients of the shear rate in both the streamwise and normal directions \citep{Jain2002}. Moreover, unlike bounded heap flow, the flowing layer thickness in rotating tumbler flow changes significantly along the length of the flowing layer. In addition, the flowing layer length changes in non-circular rotating tumblers, which can result in different particle configurations (such as radially segregated core patterns or striped patterns \citep{Hill1999}). In these cases, $\Lambda$ and $Pe$ change in both space and time.

\acknowledgements
%\section{Acknowledgements}
We thank Karl Jacob and Ben Freireich for helpful discussions. We also gratefully acknowledge financial support from The Dow Chemical Company. C.\ P.\ S.\ was supported by NSF Grant No. CMMI-1000469.

\appendix
\section{Averaging method}
\label{average}
To obtain local values of the quantities obtained from DEM simulations, the flowing layer in figure \ref{geometry}(b) is divided into non-overlapping bins of size $\Delta x=1$~cm, $\Delta y=T$, and $\Delta z=1$~mm, unless otherwise noted. The kinematic details of each particle at each time instant are obtained from DEM simulations. Based on this information, various time-averaged quantities for each bin can be calculated, as indicated below.

\textit{Solids volume fraction and volume concentration}: In each bin the solids volume fraction of each species $i$ averaged over $N$ time steps is calculated as

\begin{equation}
  f_i=\frac{1}{N} \sum_{k=1}^N \frac{\sum_j V_{ijk}}{V_{\rm{bin}}}.
  \label{f}
\end{equation}
Here, $k$ refers to time step, and $j$ labels the particle (of species $i$) that is partly or fully in the bin. $V_{ijk}$ is the fractional volume of particle $j$ at time step $k$ in the bin. $V_{\rm bin}$ = $\Delta x\Delta zT$ is the total volume of the bin. Thus, the total solids volume fraction is $f=\sum_i f_i$, and the volume concentration of species $i$ is $c_i=f_i/f$. Therefore, the number fraction in each bin for small particles $n_s=\frac {c_sR^3}{c_sR^3+c_l}$ and large particle $n_l=\frac {c_l}{c_sR^3+c_l}$, where $R=d_l/d_s$ is the ratio of large particle diameter $d_l$ to small particle diameter $d_s$. The local mean particle diameter is $\bar d=n_sd_s+n_ld_l$.

\textit{Mean velocity and percolation velocity}: The velocity component in the streamwise direction of species $i$, $u_i$, averaged over $N$ time steps in each bin is calculated as,

\begin{equation}
  u_i=\frac{1}{N} \sum_{k=1}^N \frac{\sum_j u_{ijk}V_{ijk}}{\sum_j V_{ijk}}.
  \label{u}
\end{equation}
Here, $u_{ijk}$ is the velocity component in the streamwise direction of particle $j$ at time step $k$ in the bin. The time-averaged velocity components in other directions including $v_i$ and $w_i$, as well as for the mixture ($u,v,w$), are calculated similarly. The percolation velocity for species $i$ in each bin is calculated as $w_{p,i}=w_i-w$.

\textit{Diffusion Coefficient}: The diffusion coefficient, $D$, of the mixture is calculated only in the normal direction. The time evolution of the non-affine part of trajectory is tracked by calculating the mean squared displacement, $\left< \Delta Z(\Delta t)^2\right>$, where $\Delta Z(\Delta t)=z(t_0+\Delta t)-z(t_0)-\int_{t_0}^{t_0+\Delta t} w(t)dt$ for each individual particle in each bin \citep{besseling2007,Wandersman2012}. Here, $w(t)$ is the local mean normal velocity at $t$, and $\left<*\right>$ denotes the ensemble average. The diffusion coefficient is then calculated based on $\left<\Delta Z^2\right>=2D\Delta t$ \citep{Utter2004}.

\section{Shear rate-dependent diffusion coefficient}
\label{sensitivity}

\begin{figure}
  \centerline{\includegraphics[scale=0.5]{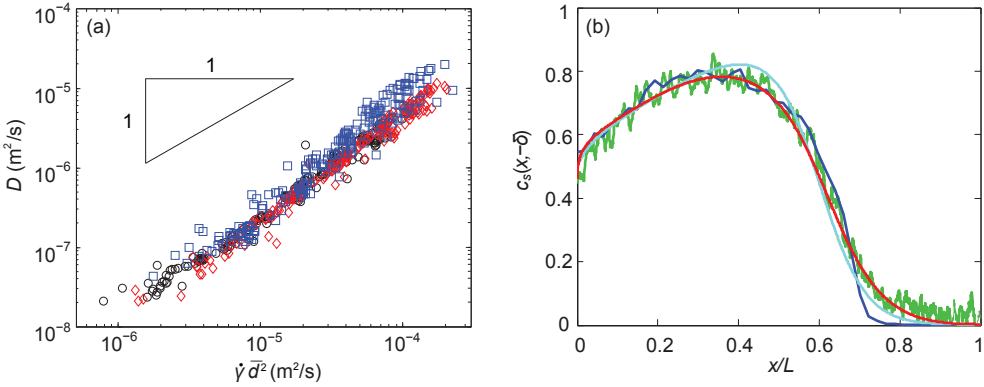}}% Images in 100% size
  \caption{(a) Log-log plot of diffusion coefficient vs.\ ${\dot \gamma} {\bar d}^2$ in the flowing layer for three simulation runs: 1 and 2~mm particles at $Q=4.57\times 10^3$~mm$^3$/s (black circles), 1 and 2~mm particles at $Q=1.52\times 10^4$~mm$^3$/s (red diamonds), and 1 and 3~mm particles at $Q=1.52\times 10^4$~mm$^3$/s (blue squares). (b) The effects of a spatially varying diffusion coefficient on theoretical prediction. Theoretical predictions with a spatially varying $D$ [dark grey (red online)] more closely match experiment [lighter grey (green online)] and simulation [black (blue online)] than predictions with constant $D$ [lightest grey (cyan online)] for 1 and 2~mm diameter particles at $Q=4.57\times 10^3$~mm$^3$/s. $\Lambda=0.78$ and $Pe$ = 19 for the constant diffusion coefficient.}
\label{D}
\end{figure}

While theoretical predictions based on a constant diffusion coefficient measured from DEM simulations accurately predict segregation as shown in figure~\ref{fig:Compare_sim_theory}, the diffusion coefficient actually depends on the shear rate (equation~\eqref{eq:diffusion}). As shown in figure~\ref{D}(a), $D\sim\dot{\gamma}\bar d^2$. Using the exponential velocity profile in equation~\eqref{w_z_exp}, $\dot{\gamma}\sim (1-x/L)\text{exp}(kz/\delta)$. Therefore, the expression for the spatially varying diffusion coefficient is
\begin{equation}
D=D_{m}(1-x/L)\text{exp}(kz/\delta),
\label{eq:D}
\end{equation}
where $D_m$ is the maximum diffusion coefficient at $(x,z)=(0,0)$ and can be measured from DEM simulations.

Figure~\ref{D}(b) shows the theoretical prediction of small particle concentration at steady state based on both a constant diffusion coefficient $D_{\rm mean}$ (measured from simulations) and a spatially varying diffusion coefficient. Using a spatially varying diffusion coefficient results in slightly better agreement of the theoretical prediction with simulation and experiment compared to the prediction based on constant $D$, though the difference is not large. In the case of the spatially varying diffusion coefficient, the diffusive fluxes in the upstream region are larger than in the downstream region so that more small particles remain in the flowing layer and are advected to the downstream region of the heap. However, the prediction based on constant $D$ matches both simulation and experiment quite well, indicating that neglecting the dependence of the diffusion coefficient on spatially varying shear rate is a reasonable approximation.

\section{Time scales}
\label{time_scales}

Here, we justify the time-scales in equation~\eqref{eq:timescales}. Although we show in \S \ref{comparison} that an exponential velocity profile better predicts particle configurations, for the purpose of this analysis, a linear profile in equation~\eqref{w_z_linear} is sufficient.

The advection time $\tilde{t}_{a}$ is the median time a particle spends in the flowing layer. A particle entering the flowing layer at $(\tilde{x},\tilde{z})=(0,1/\sqrt{2}-1)$ exits the flowing layer at $(1/2,-1)$. This implies that the advection time is the time it takes a particle starting at $(\tilde{x},\tilde{z})=(0,1/\sqrt{2}-1)$ to exit the flowing layer since half the particles fall out of the flowing layer sooner ($\tilde{x}<0.5$). Equation~\eqref{w_z_linear} yields $\tilde{t}_{a}=\int_{1/\sqrt{2}-1}^{-1}(1/\tilde{w})d\tilde{z}\approx2$.

The segregation time $\tilde{t}_{s}$ is given by the time it takes a small particle to percolate through a matrix of large particles for half the flowing layer depth. Again assuming a linear velocity profile, equations~\eqref{w_z_linear} and \eqref{percolation_v} give $w_{p,i}=(2Sq/\delta^{2})(1-x/L)$. Therefore,
\begin{equation}
\tilde{t}_{s}=\frac{\delta/2}{w_{p,i}}=\frac{\delta^{3}}{2Sq(1-x/L)}.
\end{equation}
Nondimensionalizing the above expression and taking $x=0$ (as segregation upstream is more important for determining particle configurations) yields
\begin{equation}
\tilde{t}_{s}=\frac{0.5}{\Lambda}.
\label{ts}
\end{equation}

To determine the diffusion time $\tilde{t}_{d}$, consider equation~\eqref{eq:ND} with $\bm{u}=\bm{0}$ and $\Lambda=0$,
\begin{equation}
\frac{\partial c}{\partial \tilde{t}}=\frac{1}{Pe}\frac{\partial^{2}c}{\partial\tilde{z}^{2}}
\label{td}
\end{equation}
on $-1\le \tilde{z} \le0$ with no flux boundary conditions and arbitrary initial condition. These conditions are chosen to match the original problem as closely as possible in the absence of advection and segregation. The solution to equation~\eqref{td} is
\begin{equation}
c=a_{0}+\sum_{n=1}^{\infty}a_{n}\text{cos}(n\pi\tilde{z})\exp\left[-\left(\frac{n\pi}{Pe}\right)^{2}\tilde{t}\right].
\end{equation}
We set the diffusion time to the time it takes the dominant ($n=1$) mode to decay by a factor of $1/e$, giving $\tilde{t}_{d}={Pe}/\pi^{2}\approx {Pe}/10$.

\bibliographystyle{jfm}
%\bibliography{heap_model}

\end{document}